\documentclass[]{article}  

\usepackage{graphicx,amsfonts,amssymb,latexsym,url}
\newcommand{\lemlab}[1]{\label{lemma:#1}}
\newcommand{\theolab}[1]{\label{theo:#1}}

\newcommand{\figlab}[1]{\label{fig:#1}}
\newcommand{\seclab}[1]{\label{section:#1}}

\newcommand{\lemref}[1]{\ref{lemma:#1}}

\newcommand{\theoref}[1]{\ref{theo:#1}}

\newcommand{\figref}[1]{\ref{fig:#1}}
\newcommand{\eqref}[1]{(\ref{eq:#1})}
\newcommand{\secref}[1]{\ref{section:#1}}

\newtheorem{theorem}{Theorem}
\newtheorem{lemma}[theorem]{Lemma}
\newtheorem{cor}[theorem]{Corollary}
\newtheorem{conj}{Conjecture}[section]
{\catcode`\@=11
\gdef\setft#1#2#3{%
\def\@oddfoot{
{\setbox0=\hbox{#1}
\setbox1=\hbox{#3}
\ifdim\wd0>\wd1
\dimen0=\wd0
\box0\hfil#2\hfil\hbox to\dimen0{\hfil\hfil\box1}
\else \dimen0=\wd1
\hbox to\dimen0{\box0\hfil }\hfil#2\hfil\box1 \fi
}}} }

\def\complaint#1{}
\def\withcomplaints{
\newcounter{mycomplaints}
\def\complaint##1{\refstepcounter{mycomplaints}%
\ifhmode%
\unskip%
{\dimen1=\baselineskip \divide\dimen1 by 2 %
\raise\dimen1\llap{\tiny -\themycomplaints-}}\fi%
\marginpar{\tiny [\themycomplaints]: ##1}}%
}

\def\M{{\cal M}}
\def\R{\mathbb{R}}
\def\S{\mathbb{S}}

\newlength\abovesectionskip
\newlength\belowsectionskip
\makeatletter
\def\section{\@startsection{section}{1}{\z@}{-\abovesectionskip}%
               {\belowsectionskip}{\normalfont\Large\bfseries}}
\makeatother

\makeatletter
\@ifundefined{abovecaptionskip}{\newlength\abovecaptionskip}
\long\def\@makecaption#1#2{
   \vskip \abovecaptionskip
   \setbox\@tempboxa\hbox{{\sf\footnotesize \textbf{#1.} #2}}
   \ifdim \wd\@tempboxa >\hsize         % IF longer than one line:
       {\sf\footnotesize \textbf{#1.} #2\par}% THEN set as ordinary paragraph.
     \else                              %   ELSE  center.
       \hbox to\hsize{\hfil\box\@tempboxa\hfil}
   \fi}
\def\@opargbegintheorem#1#2#3{\trivlist
	\item[\hskip\labelsep{\bf\boldmath #1\ #2\ (#3).}]\sl}
\def\@begintheorem#1#2{\trivlist
	\item[\hskip\labelsep{\bf\boldmath #1\ #2.}]\sl}

\def\newproof#1#2{\@ifnextchar({\@snproof{#1}{#2}}{\@snproof{#1}{#2}(\rm)}}
\def\@snproof#1#2(#3){\@ifnextchar[{\@xnproof{#1}{#2}{#3}}
                                   {\@xnproof{#1}{#2}{#3}[]}}

\def\@xnproof#1#2#3[#4]{%\expandafter\@ifdefinable\csname #1\endcsname
{\global\@namedef{#1}{\@prf{#2}{#3}}\global\@namedef{end#1}{\@endprf{#4}}}}

\def\@prf#1#2{\@ifnextchar[{\@xprf{#1}{#2}}{\@yprf{#1}{#2}}}
\def\@xprf#1#2[#3]{\@yprf{#1\ (#3)}{#2}}
\def\@yprf#1#2{\begin{trivlist}\item[\hskip\labelsep{\bf\boldmath #1:}]#2}

\def\@endprf#1{#1\end{trivlist}}

\@ifundefined{square}{\let\square\Box}{}	% grumble latex2e 
\def\QED{\ensuremath{{\square}}}
\def\markatright#1{\leavevmode\unskip\nobreak\quad\hspace*{\fill}{#1}}
\def\qed{\markatright{\QED}}
\newproof{proof}{Proof}[\qed]
\newproof{sketch}{Proof Sketch}[\qed]
\makeatother

\def\Real{\mathrm{I\!R}}
\def\set#1{\lbrace #1 \rbrace}

\begin{document}

\title{\textbf{Vertex-Unfoldings\\of Simplicial Manifolds}}

\author{
Erik~D.~Demaine%
\thanks{MIT Laboratory for Computer Science,
	200 Technology Square, Cambridge, MA 02139, USA.
	edemaine@mit.edu.}  %JOR: Need Erik's grant info
\and
David~Eppstein%
\thanks{Department of Information and Computer Science, 
	University of California, Irvine CA
	92697-3425, USA.  
	eppstein@ics.uci.edu.
	Supported by NSF grant CCR-9912338.}
\and
Jeff~Erickson%
\thanks{Department of Computer Science, University of Il\-linois at
	Ur\-bana-Cham\-paign; 
	http://\allowbreak www.cs.uiuc.edu/\allowbreak\~{}jeffe; jeffe@cs.uiuc.edu.
	Partially supported by a Sloan Fellowship and NSF CAREER award
	CCR-0093348.}
\and
George~W.~Hart%
\thanks{http://www.georgehart.com/;
	george@georgehart.com.}
\and
Joseph~O'Rourke%
\thanks{
	Department of Computer Science, Smith Col\-lege, North\-ampton,
	MA 01063, USA.
	orourke@\allowbreak cs.smith.edu.
	%Supported by NSF grant CCR-9731804.
	Supported by NSF Distinguished Teaching Scholars award
        DUE-0123154.
}
}

\date{}  %Comment out so date DOES appear during draft stage.

\maketitle

\begin{abstract}
We present an algorithm to unfold any triangulated $2$-manifold (in
particular, any simplicial polyhedron) into a non-overlapping,
connected planar layout in linear time.  The manifold is cut only
along its edges.  The resulting layout is connected, but it may have a
disconnected interior; the triangles are connected at vertices, but
not necessarily joined along edges.  We extend our algorithm to
establish a similar result for simplicial manifolds of arbitrary
dimension.
\end{abstract}


% ----------------------------------------------------------------------
\section{Introduction}
\seclab{Introduction} 

%\paragraph{Background.}
It is a long-standing open problem to determine whether every convex
polyhedron can be cut along its edges and unfolded flat in one piece
without overlap, that is, into a simple polygon.  This type of
unfolding has been termed an \emph{edge-unfolding}; the unfolding
consists of the facets of the polyhedron joined along edges.  In
contrast, unfolding via arbitrary cuts easily leads to nonoverlap.
O'Rourke~\cite{o-fucg-00} gives a history of the edge-unfolding
problem and its applications to manufacturing.  Recently it was
established that not every nonconvex polyhedron can be edge-unfolded,
even if the polyhedron is \emph{simplicial}, that is, all of its faces
are triangles~\cite{bdek-up-99,bdems-uptf-01}.

In this paper we loosen the meaning of ``in one piece'' to permit a
non-overlapping connected region that (in general) does not form a
simple polygon, because its interior is disconnected.  We call such an
unfolding a \emph{vertex-unfolding}: facets of the polyhedron are
connected in the unfolding at common vertices, but not necessarily
along edges.  With this easier goal we obtain a positive result:
%the surface of every simplicial polyhedron,
%convex or nonconvex, of any genus, may be cut along edges and unfolded
%to a planar, non-overlapping, connected layout.  

%\begin{theorem}
%Any connected triangulated $2$-manifold can be vertex-unfolded: cut
%along edges and unfolded to a non-overlapping, connected planar
%layout.  Moreover, a vertex-unfolding can be computed in linear time.
%\theolab{manifold.2D}
%\end{theorem}

\begin{theorem}
Every connected triangulated $2$-manifold (possibly with boundary) has
a vertex-unfolding, which can be computed in linear time.
\theolab{manifold.2D}
\end{theorem}

\noindent
This result includes simplicial polyhedra of any genus, manifolds with
any number of boundary components, and even manifolds like the Klein
bottle that cannot be topologically embedded in $3$-space.  Our proof
relies crucially on the restriction that every face is a triangle.
The problem remains open for nonsimplicial polyhedra with simply
connected or even convex faces; see Section~\secref{Discussion}.

We extend this result in the natural way to higher dimensions in
Section~\secref{Higher.Dimensions}.


% ----------------------------------------------------------------------
\section{Algorithm Overview}
\seclab{Overview}

Let $\M$ be a triangulated $2$-manifold, possibly with boundary.
Following polyhedral terminology, we refer to the triangles of $\M$ as
\emph{facets}.  The \emph{(vertex-facet) incidence graph} of $\M$ is
the bipartite graph whose nodes are the facets and vertices of $\M$,
with an arc $(v,f)$ whenever $v$ is a vertex of facet~$f$.  A
\emph{facet path} is a trail $(v_0,f_1,v_1,f_2,v_2,\dots,f_k,v_k)$ in
the incidence graph of $\M$ that includes each facet node exactly
once, but may repeat vertex nodes.  In any facet path, $v_{i-1}$ and
$v_i$ are distinct vertices of facet $f_i$ for all $i$.  Because each
facet node appears only once, no arc is repeated.  A \emph{facet
cycle} is a facet path that is also a circuit, that is, where $v_0 =
v_k$.

Our algorithm relies on this simple observation:

\begin{lemma}
If $\M$ has a facet path, then $\M$ has a vertex-unfolding in which
each triangle of the path occupies an otherwise empty vertical strip
of the plane.  \lemlab{strip}
\end{lemma}

\begin{proof}
Let $p$ be a facet path of $\M$.  Suppose inductively that a facet
path $p$ has been laid out in strips up to facet $f_{i-1}$, with all
triangles left of vertex $v_i$, the rightmost vertex of $f_{i-1}$.
Let $(v_i, f_i, v_{i+1})$ be the next few nodes in $p$; recall that
$v_i \neq v_{i+1}$.  Rotate facet $f_i$ about vertex $v_i$ so that
$v_i$ is leftmost and $v_{i+1}$ rightmost, and the third vertex of
$f_i$ lies horizontally between.  Such rotations exist because $f_i$
is a triangle.  Place $f_i$ in a vertical strip with $v_i$ and
$v_{i+1}$ on its left and right boundaries.  Repeating this process
for all facets in $p$ produces a non-overlapping vertex-unfolding.
\end{proof}

Thus to prove Theorem~\theoref{manifold.2D} it suffices to prove that
every connected triangulated $2$-manifold has a facet path.
In addition, the existence of facet paths has two applications:
%consequences:

  \begin{enumerate}
  \item
    The vertex-unfolding resulting from Lemma~\lemref{strip} can be viewed as
    a \emph{hinged dissection}~\cite{f-dpf-97} of the surface.  Thus
    we demonstrate a hinged dissection for any triangulated $2$-manifold.
  \item
    A facet path also yields an
    ``ideal rendering'' of any triangulated surface on a computer graphics
    system with a $1$-vertex cache: each triangle shares one vertex with the
    previous triangle in the graphics pipeline.  This result is in some sense
    best possible: an ideal rendering for a 2-vertex cache in which every
    adjacent pair of triangles shares two vertices is not always achievable,
    because there are triangulations whose dual graphs have
    no Hamiltonian path~\cite{ahms-htfr-96}.
    %\cite{xhm-fesps-99}
  \end{enumerate}

It might be more pleasing to obtain a vertex-unfolding based on
a \emph{noncrossing facet path}, one that does not include a pattern
$(\ldots, A, v, C, \ldots, B, v, D, \ldots)$
with the facets incident to the vertex $v$ appearing in the cyclic order
$A, B, C, D$.
Because a facet path has either no or at most two odd nodes (its endpoints),
and because any such planar graph has a noncrossing Eulerian trail,
we can convert any facet path into a noncrossing facet path.
%We also point out that all vertex-unfoldings (in particular those from
%Lemma~\lemref{strip} and Theorem~\theoref{manifold.2D}) can be converted into
%vertex-unfoldings with the desirable property that the connections between
%the~$d$ separated copies of a vertex incident to~$d$ facets form a noncrossing
%pattern.
Specifically, we can replace each crossing pair of vertex-to-vertex connections
with one of the two alternate pair of connections, whichever
alternate pair keeps the vertex-unfolding connected.  See
Figure~\figref{undo-cross}.

\begin{figure}[htbp]
\centering
\includegraphics[width=0.98\linewidth]{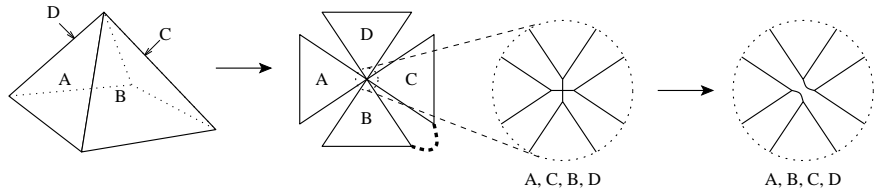}
\caption{Vertex-unfolding the top four triangles of the
         regular octahedron and making the connections planar.}
\figlab{undo-cross}
\end{figure}

Figure~\figref{cube} shows a vertex-unfolding of the triangulated
surface of a cube, obtained from a facet path by our algorithm.
Figure~\figref{vunf.all} shows several more complex examples,
generated from random convex polyhedra using an earlier, less general
version~\cite{deeho-vusp-01} of the algorithm we present in the next
section.  In our examples, we permit the triangles to touch along
segments at the strip boundaries, but this could easily be avoided if
desired so that each strip boundary contains just the one vertex
shared between the adjacent triangles.

%%%%%%%%%%%%%%%%%%%%%%%%%%%%%%%%%Figure Begin
\begin{figure}[htbp]
\centering
\includegraphics[width=0.7\linewidth]{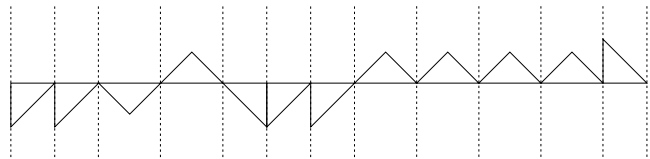}
\caption{Laying out a facet path in vertical strips.}
\figlab{cube}
\end{figure}
%%%%%%%%%%%%%%%%%%%%%%%%%%%%%%%%%Figure End

%%%%%%%%%%%%%%%%%%%%%%%%%%%%%%%%%Figure Begin
\begin{figure}[htbp]
\centering
\includegraphics[width=\linewidth]{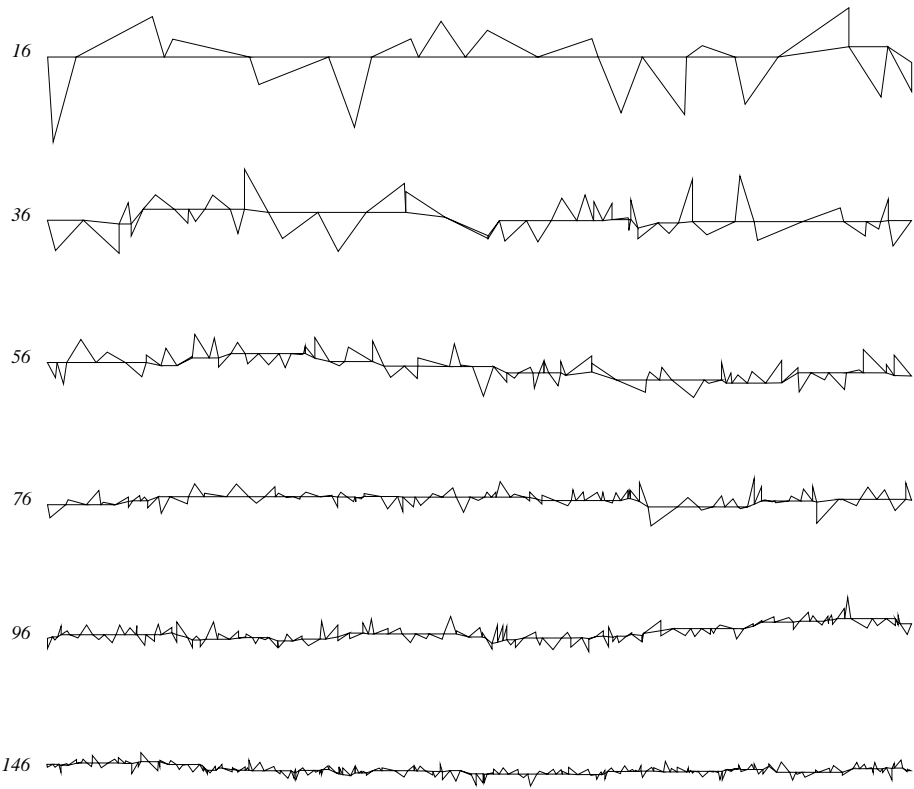}
\caption{Vertex-unfoldings of random convex polyhedra (generated by
code from O'Rourke~\cite{o-cgc-98}).  The number of triangles is
indicated to the left of each unfolding.}  \figlab{vunf.all}
\end{figure}
%%%%%%%%%%%%%%%%%%%%%%%%%%%%%%%%%Figure End


% ----------------------------------------------------------------------
\section{Facet Paths in 2-Manifolds}
\seclab{newcycle}

In this section, we prove that every triangulated $2$-manifold has a
facet path, which, by Lemma~\lemref{strip}, yields
Theorem~\theoref{manifold.2D}.

A \emph{$d$-manifold} is a topological space such that every point has
a neighborhood homeomorphic to the neighborhood of some point in a
closed halfspace in $\R^d$.  Interior points of a $2$-manifold have
neighborhoods homeomorphic to a disk, and boundary points have
neighborhoods homeomorphic to a half-disk.  (Thus, the
vertex-unfoldings in Figures~\figref{cube} and~\figref{vunf.all} are
not manifolds.)  A $2$-manifold constructed from (topological)
triangles is called a \emph{triangulated $2$-manifold}.

The \emph{weak dual} of a triangulated $2$-manifold $\M$ is a graph
$\M^*$ with a node for every triangle and an arc between any pair of
triangles that share a common edge.  Let $T^*$ be an arbitrary
spanning tree of this dual graph.  If we glue the triangles of $\M$
together according to the edges in $T^*$, we obtain a
\emph{topological unfolding} $T$, a simplicial complex with the
topology of a triangulated polygon with no interior vertices.  (If we
are lucky, we can obtain a \emph{geometric} unfolding of $\M$ by
embedding $T$ in the plane, but as mentioned earlier, this is not
always possible.)  Any facet path or facet cycle of $T$ can be mapped
to a facet path or facet cycle of $\M$; recall that a facet path may
repeat vertex nodes.

\begin{figure}[htb]
\centerline{\footnotesize\sf
\begin{tabular}{cc}
	\includegraphics[scale=0.5]{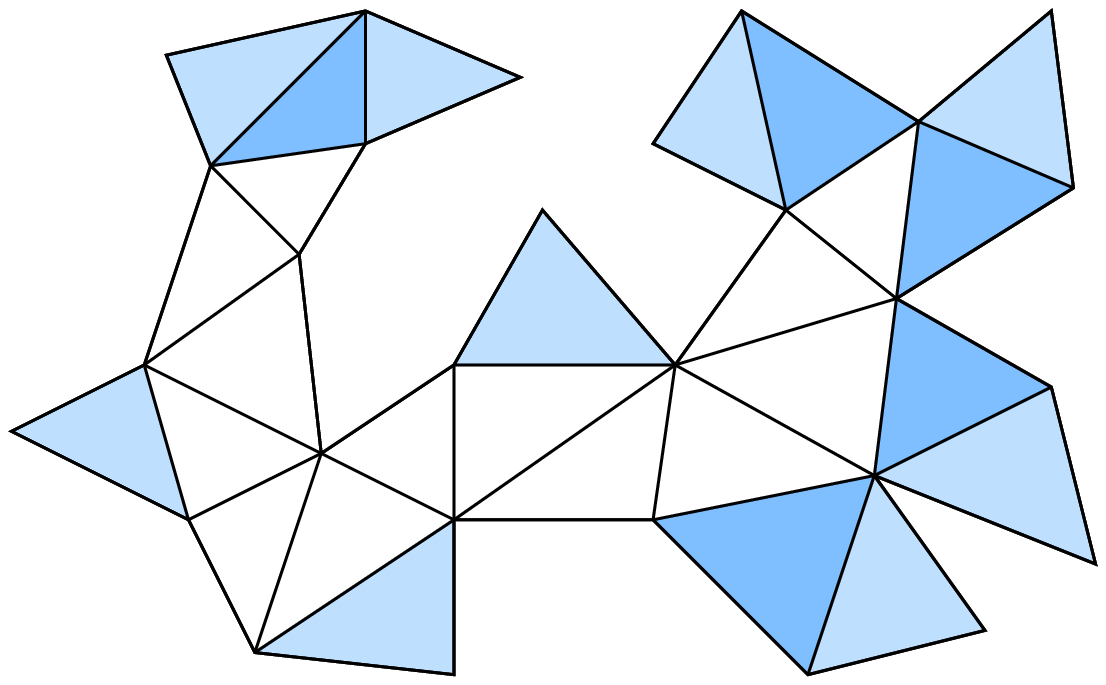} &
	\includegraphics[scale=0.5]{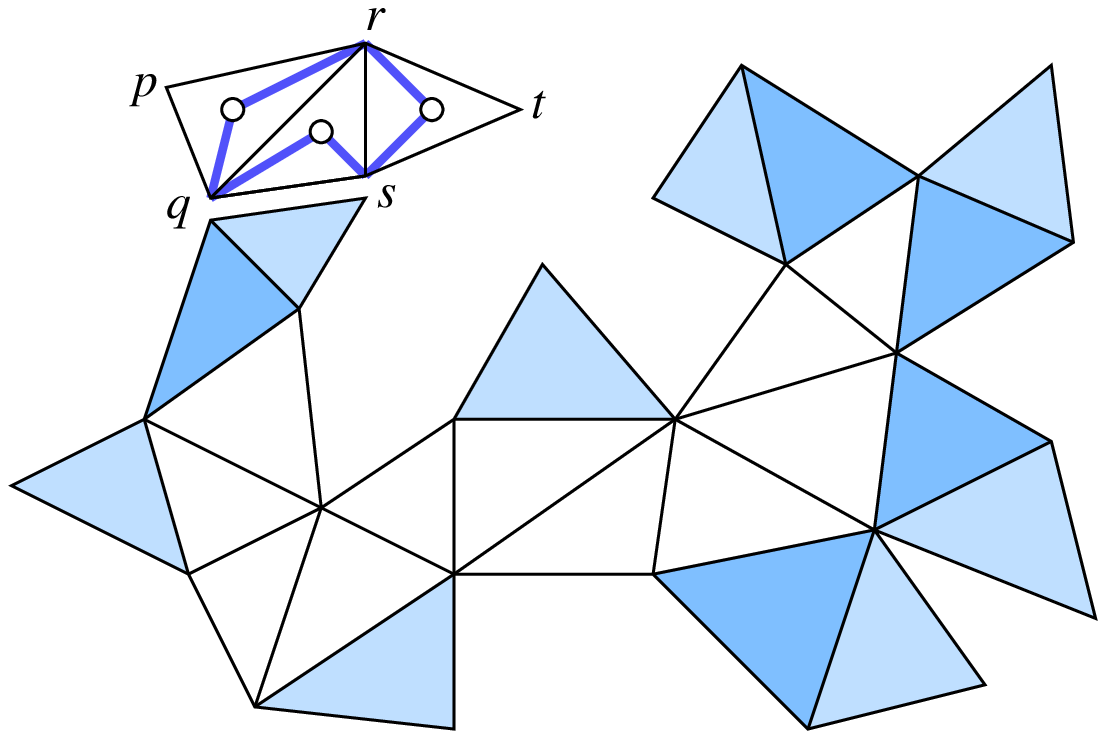} \\
	(a) & (b)
\\[2ex]
	\includegraphics[scale=0.5]{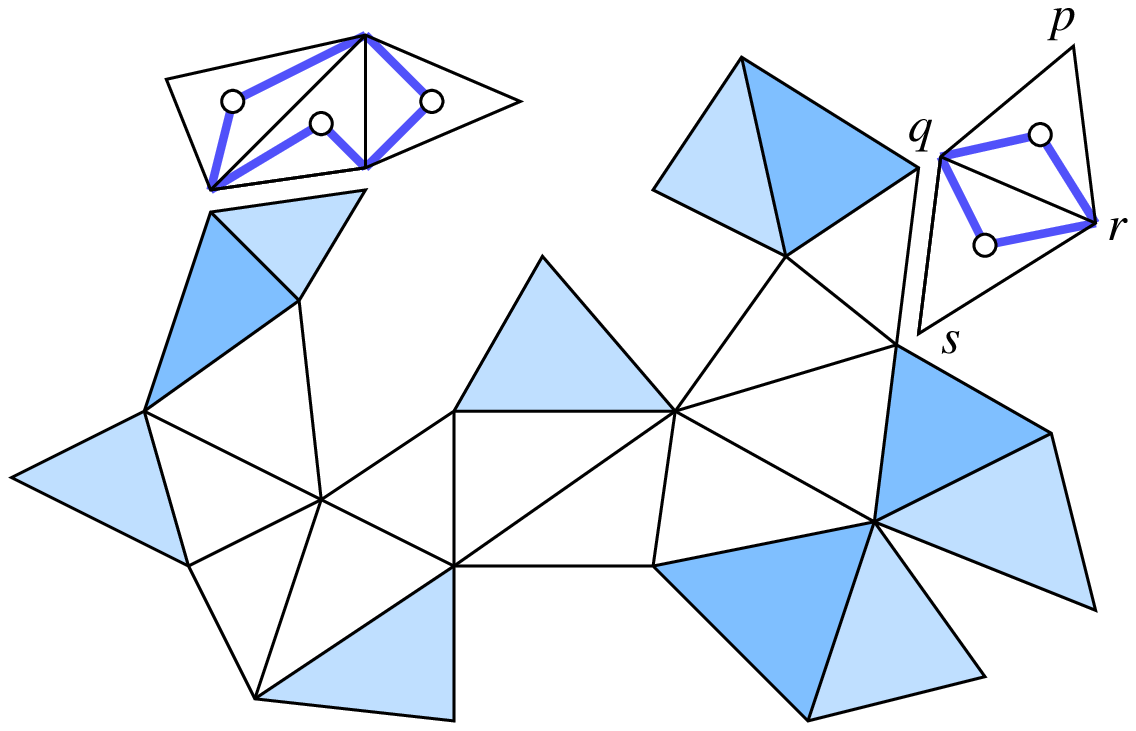} &
	\includegraphics[scale=0.5]{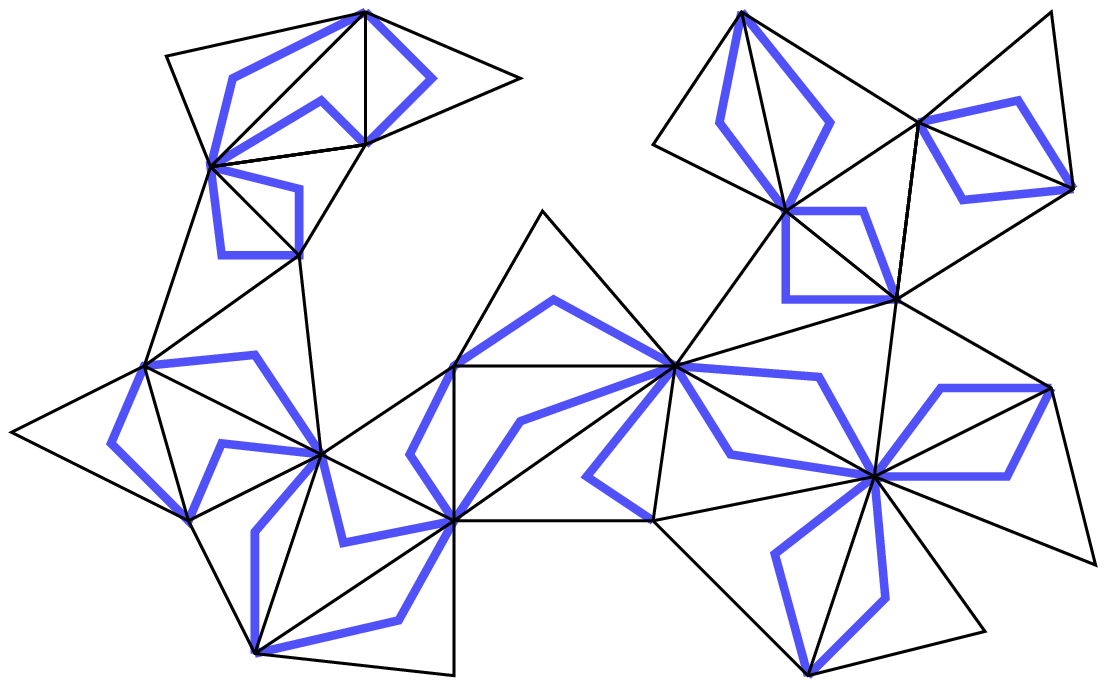} \\
	(c) & (d)
\end{tabular}}
\caption{(a) A polygon triangulation.  Lightly shaded triangles are
ears; darker shaded triangles are hats.  (b) Removing a Mickey House
hat.  (c) Removing a dunce cap.  (d) A scaffold produced by our
algorithm.}
\label{F:scaffold}
\end{figure}

A \emph{scaffold} is a subgraph of the incidence graph in which every
facet appears and has degree~$2$ and at most two vertices have odd degree;
if every vertex has even degree, we call it an \emph{even} scaffold.
(See Figure \ref{F:scaffold}(d) below for an example of a scaffold.)
The edges of any facet path form a scaffold, and any Euler walk
through a connected scaffold gives us a facet path.  Thus, our goal is
to find a connected scaffold for~$T$.

First we establish a slightly weaker result:

\begin{lemma}
\label{L:scaffold}
Every triangulated polygon with no interior vertices has a (possibly
disconnected) scaffold.
\end{lemma}

\begin{proof}
Let $T$ be a triangulated polygon with no interior vertices.  We prove the
lemma by induction on the number of triangles, with two base cases.  If $T$ is
empty, we are done.  If $T$ is a single triangle, then a path between any two
vertices is a scaffold.  Henceforth, assume that $T$ has at least two
triangles.

An \emph{ear} in $T$ is a triangle that is adjacent to at
most one other triangle.  We call a triangle in $T$ a \emph{hat} if it
is adjacent to at least one ear and at most one non-ear.  If we remove
all the ears from $T$, we obtain a new triangulated polygon $T'$.  See
Figure~\ref{F:scaffold}(a).  If $T'$ is nonempty, then it has at least
one ear, and every ear in $T'$ is a hat in $T$.  On the other hand, if
$T'$ is empty, then $T$ consists of exactly two triangles, which are
both ears and hats.  In either case, $T$ contains at least one hat.

To perform the induction, we choose a hat in $T$, find a cycle in the
facet-vertex of that hat and (at most two of) its adjacent ears, and
recursively construct a scaffold for the remaining triangulation.  We
have two inductive cases.

Suppose $T$ has a hat $H = qrs$ with at least two ears $E = pqr$ and
$F = rst$ (a `Mickey Mouse hat').  See Figure \ref{F:scaffold}(b).  We
construct a cycle $(r,E,q,H,s,F,r)$ through the facet-vertex of these
three triangles, and recursively construct a scaffold for the smaller
triangulation $T \setminus \set{H,E,F}$.

Otherwise, let $H = qrs$ be a hat with just one adjacent ear $E = pqr$
(a `dunce cap').  See Figure \ref{F:scaffold}(c).  We construct a
cycle $(q,H,r,E,q)$ through the facet-vertex of those two triangles,
and recursively construct a scaffold for the smaller triangulation $T
\setminus \set{H,E}$.
\end{proof}

Following the proof gives us an easy linear-time algorithm, consisting
of a simple depth-first traversal of the input triangulation's dual
tree.  Figure \ref{F:scaffold}(d) shows a scaffold computed by our
algorithm.  The scaffold we construct may be disconnected, but we
show how to make it connected using a series of local operations.

\begin{figure}[htb]
\centerline{\includegraphics[height=1.25in]{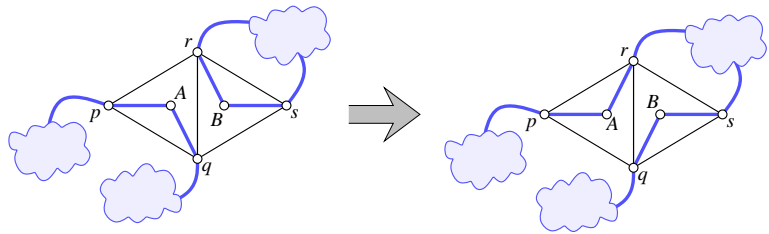}}
\caption{Joining two components of a scaffold with a flip.  Clouds
hide the rest of the components.}
\label{F:flip}
\end{figure}

Let $S$ be a scaffold with more than one component.  If $S$ is not
actually an even scaffold, its two odd vertices must be in the same
component; every other component has an Euler circuit and so must be
$2$-connected.  Choose a pair of triangles $A = pqr$ and $B = qrs$
that lie in different components of $S$.  Without loss of generality,
suppose $S$ contains the edges $(p,A)$, $(q,A)$, $(r,B)$, and $(s,B)$.
At most one of the edges $(q,A)$ and $(r,B)$ is a bridge
(its removal would disconnect the graph).  If we
remove edges $(q,A)$ and $(r,B)$ and add edges $(r,A)$ and $(q,B)$, we
obtain another scaffold $S'$ with one fewer component than $S$.  See
Figure \ref{F:flip}.  Repeating this process for each adjacent pair of
components gives us a connected scaffold.

\begin{theorem}
Every connected triangulated $2$-manifold (possibly with boundary)
has a facet path.
\theolab{2-manif.path}
\end{theorem}

\subsection{Checkered Triangulations and Facet Cycles}

We can strengthen this result slightly, by showing that most properly
triangulated $2$-manifolds actually have a facet \emph{cycle}.  This
permits the strip layout of Lemma~\lemref{strip} to start on the left
with any given triangle of the manifold.

We call a polygon triangulation \emph{checkered} if there is a 
$2$-coloring of the triangles so that every white triangle has three
(necessarily black) neighbors.  See Figure~\ref{F:layered}(a).

\begin{lemma}
\label{L:layered}
A polygon triangulation with no interior vertices has a facet cycle if and only
if it is not checkered.
\end{lemma}

\begin{proof}
Let $T$ be a triangulated polygon with no interior vertices.
First we prove by induction that no checkered triangulation has a
facet cycle.  The base case is a single (black) triangle, which
clearly has no facet cycle.  In any other checkered triangulation $T$,
we can always find a Mickey Mouse hat: two black ears adjacent to a
common (white) triangle.  If $T$ has a facet cycle, it must contain a
subcycle of six edges inside the Mickey Mouse hat and another subcycle
through the rest of the triangulation.  But the rest of the
triangulation is checkered, so by the induction hypothesis, it has no
facet cycle.  See Figure~\ref{F:layered}(b).

\begin{figure}[htb]
\centering\footnotesize\sf
\begin{tabular}{c}
	\includegraphics[height=1.5in]{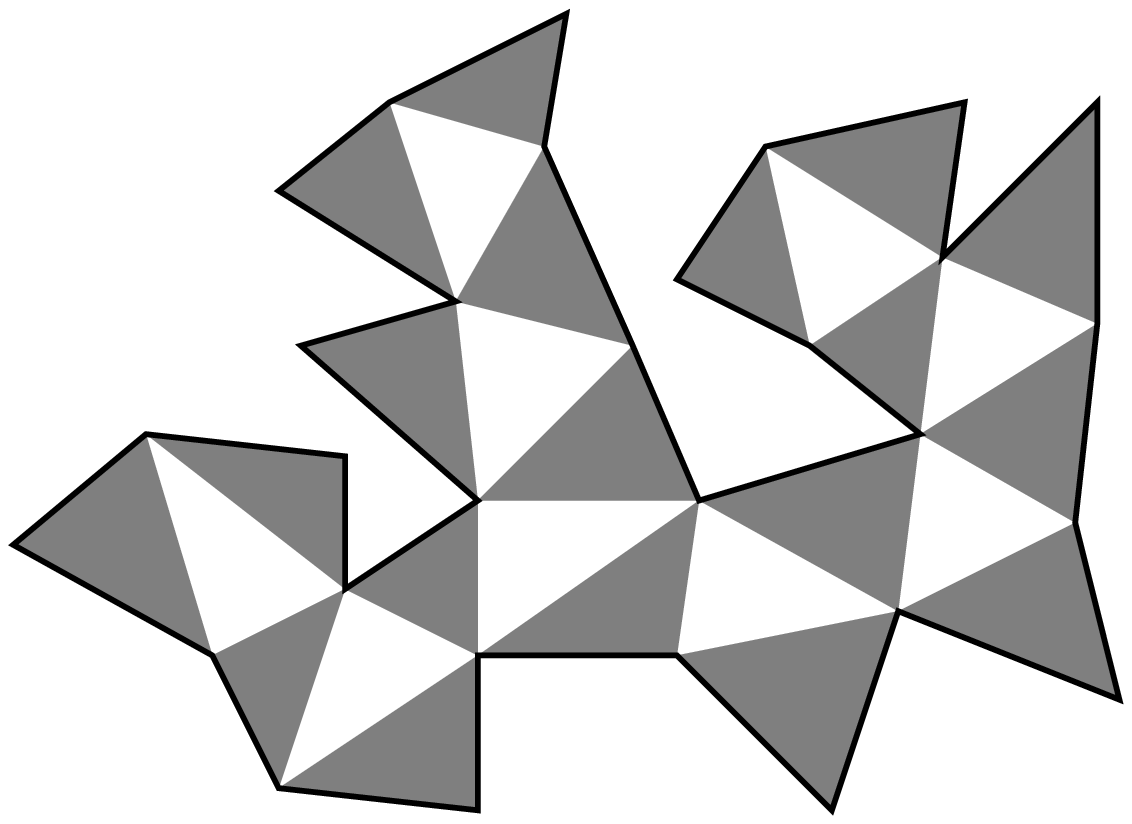} \\ (a)
\end{tabular}
\hfil
\begin{tabular}{c}
	\includegraphics[height=1.5in]{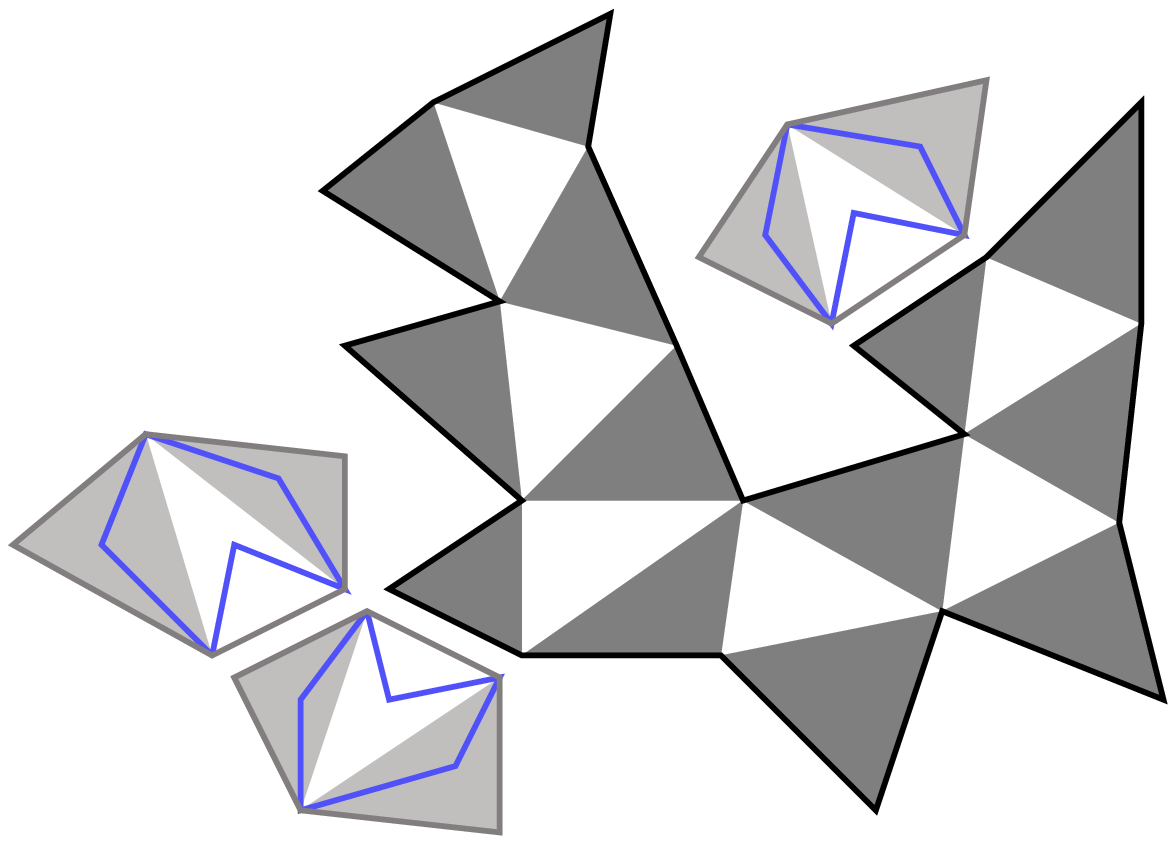} \\ (b)
\end{tabular}
\caption{(a)~A checkered polygon triangulation.  (b)~After removing
three Mickey Mouse hats.}
\label{F:layered}
\end{figure}

Now suppose $T$ is a non-checkered triangulation; in particular, $T$
has at least two triangles.  To prove the lemma, it suffices to show
that $T$ has an even scaffold.  Assume without loss of generality that
$T$ has \emph{no} Mickey Mouse hats; otherwise, we can remove them as
described above.  (This might actually eliminate every triangle in
$T$, but then we've computed an even scaffold!)  We cannot be left
with a single triangle because $T$ is not checkered.  Thus, $T$ has at
least two `dunce caps': hats with only one adjacent ear.  If we follow
the algorithm in Lemma \ref{L:scaffold}, removing Mickey Mouse hats
whenever possible, the triangulation always contains at least one
dunce cap, until either the algorithm removes every triangle or there
are exactly three triangles left.  Thus, the triangulation never
consists of a single triangle, which implies that the algorithm
constructs an even scaffold.
\end{proof}

We say that a triangulated $2$-manifold is \emph{simplicial} if it is
a simplicial complex, or equivalently, if its dual graph is simple
(has no multi-edges or loops).  Every manifold constructed from
geometric triangles is simplicial.  (A non-simplicial manifold is
shown below in Figure~\ref{F:improper}.)

\begin{lemma}
Every connected simplicial $2$-manifold (possibly with boundary) has a
non-checkered topological unfolding, except a checkered polygon
triangulation.
\end{lemma}

\begin{proof}
Let $\M$ be a simplicial $2$-manifold that either is multiply
connected or has interior vertices.  Assume without loss of generality
that $\M$ has a checkered topological unfolding $T$, because otherwise
we have nothing to prove.  This immediately implies that $\M$ is not
$2$-colorable.

Color the triangles of $T$ black and white, so that adjacent triangles
have opposite colors and every boundary edge of $T$ lies on a black
triangle.  Because the dual $1$-skeleton of $\M$ is a simple graph, we
can cut $T$ into two simple polygons along some edge, and then reglue
those pieces along some other pair of edges, to obtain another
combinatorial unfolding $T'$ of $\M$.  Because $\M$ is not
$2$-colorable, we must reverse the colors of one of those pieces to
obtain a proper $2$-coloring of $T'$.  Because each piece has at least
three edges, each piece has at least one edge that is on the boundary
of both $T$ and $T'$.  It follows that $T'$ has boundary edges
adjacent to both black and white triangles, so $T'$ is not checkered.
\end{proof}

Combining the previous two lemmas, we conclude the following:

\begin{theorem}
Every connected simplicial $2$-manifold (possibly with boundary) has a
facet cycle, except a checkered polygon triangulation.
\end{theorem}

This theorem requires that we start with a simplicial complex.  There
are triangulated but non-simplicial $2$-manifolds that have no facet
cycle, like the triangulation of the sphere shown in
Figure~\ref{F:improper}.  However, even improperly triangulated
$2$-manifolds have facet paths.

\begin{figure}
\centerline{\includegraphics[height=1.75in]{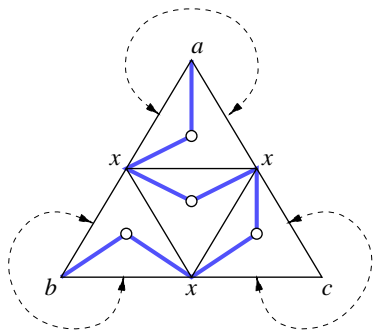}}
\caption{An improper triangulation of the sphere with a facet path but
no facet cycle.  Edges with identically labeled endpoints are
identified.}
\label{F:improper}
\end{figure}

% ----------------------------------------------------------------------
\section{Higher Dimensions}
\seclab{Higher.Dimensions}

In this final section, we generalize our results to higher dimensions.
Lemma~\lemref{strip} generalizes in the obvious way, yielding a
vertex-unfolding into parallel ``slabs'' from any facet path.  We will
show that any simplicial polyhedron has a facet cycle, and thus a
vertex-unfolding.  Like the results in the previous section, the
proofs are almost purely topological, and thus actually apply to
arbitrary triangulated manifolds, possibly with boundary, independent
of any embedding.

First let us establish some general terminology.  A \emph{simplicial
$d$-manifold} is a simplicial complex\footnote{In fact, our results
apply to \emph{pseudo-manifold $\Delta$-complexes}---sets of
$d$-simplices whose facets (our ridges) are glued together in
pairs~\cite{h-at-01}.  A pseudo-manifold $\Delta$-complex is not
necessarily a manifold, even with boundary, but it becomes a manifold
if all faces of dimension $d-2$ or less are deleted.} homeomorphic to
a $d$-dimensional manifold, possibly with boundary.  For example, any
simplicial convex $d$-polytope is a simplicial manifold homeomorphic
to the sphere $\S^{d-1}$.  Following standard polytope nomenclature,
we call the full $d$-dimensional simplices \emph{facets} and the
codimension-$1$ simplices (i.e., the simplices of dimension $d-1$)
\emph{ridges}.

The \emph{dual $1$-skeleton} $\M^*$ of a simplicial manifold $\M$ is a
simple graph, with a node for each facet and an arc between any two
facets that share a ridge.  We call a simplicial manifold $\M$
\emph{unfolded} if its dual $1$-skeleton $\M^*$ is a tree, or
equivalently, if it is conencted and its codimension-$2$ simplices all
lie on the boundary.  Every unfolded simplicial manifold is
homeomorphic to a ball.  An unfolded $3$-manifold has the topology of
a triangulated polyhedron in $\Real^3$ with no diagonals or interior
% (Steiner)
points.

An \emph{unfolding} of a connected simplicial manifold $\M$ is a pair
$(U, f)$, where $U$ is an unfolded simplicial manifold, and $f:U \to
\M$ is a simplicial map that is onto and whose restriction to the
interior of $U$ is one-to one.  The function $f$ is called the
\emph{folding map}, and the set of points in $\M$ with more than one
preimage in $U$ is called the \emph{cut set} of the unfolding.  We
will refer to $U$ as an unfolding of $\M$ when the folding map is
clear from context.  We easily observe that any spanning tree of
$\M^*$ corresponds to a unique unfolding of $\M$.  It is an open
problem, for every $d\ge 3$, whether every $d$-polytope has an
unfolding that embeds geometrically in $\Real^{d-1}$ without overlap.

The \emph{(vertex-facet) incidence graph} of a simplicial manifold has
a node for every vertex and every facet, and an arc $(v,f)$ whenever
$v$ is a vertex of facet~$f$.  A \emph{facet cycle} is a circuit in
the incidence graph that passes through every facet exactly once.
More generally, an \emph{even scaffold} is a subgraph of the incidence
graph in which every facet has degree $2$ and every vertex has even
degree.  Any Euler tour of a connected even scaffold is a facet cycle.

\begin{lemma}
\label{L:scaffold3}
For all $d\ge 3$, every unfolded simplicial $d$-manifold has an even
scaffold, except a single $d$-simplex.
\end{lemma}

\begin{proof}
Let $U$ be an unfolded simplicial $d$-manifold with $d\ge 3$ and with
more than one facet.  An \emph{ear} of $U$ is a facet that is adjacent
to only one other facet.  A \emph{hat} is a facet that is adjacent to
at least one ear and at most one non-ear.  Just as in the
two-dimensional case, every unfolded simplicial $d$-manifold with more
than one facet has at least two ears and at least one hat (and with
only $d-1$ exceptions, at least two hats).

We prove the lemma by induction.  If $U$ is non-empty, we identify a
small collection of simplices in $U$, find a cycle in the incidence
graph of those simplices, and recursively construct an even scaffold
for the remaining complex, which is still an unfolded manifold.  It is
fairly easy to construct an even scaffold for any subcomplex
consisting of a hat and its ears, and to decompose any unfolded
complex into a sequence of such complexes by removing a hat (and its
ears) and recursing~\cite{s-chcb-58}.  To keep things simple,
however, we will consider only the four following cases.

\begin{enumerate}
\item
If $U$ is empty, there is nothing to do.

\item
Suppose $U$ consists of exactly three simplices: a hat $H$ and two
ears $E$ and~$F$.  $E$ and $F$ share at least one vertex $p$; $E$ and
$H$ share at least one vertex $q\ne p$; and $H$ and $F$ share at least
one vertex $r\ne p,q$.  (In fact, we have $d-2$ choices for each of
these three vertices.)  Then $(p,E,q,H,r,F,p)$ is a cycle in the
incidence graph of $U$.

\item
Suppose some hat $H$ is adjacent to just one ear $E$.  $H$ and $E$
share an edge $pq$.  (In fact, they share an entire ridge.)  We
recursively construct an even scaffold for the subcomplex $U \setminus
\set{H,E}$, and add the cycle $(p,H,q,E,p)$.
    
\item
Finally, suppose some hat $H$ is adjacent to more than one ear and $U$
has more than three facets.  Let $E$ and $F$ be any two ears adjacent
to $H$.  $E$ and $F$ share an edge $pq$.  (In fact, they share an
entire face of dimension $d-2 > 0$; this is the only step of the proof
that requires $d\ge 3$.)  We recursively construct an even scaffold
for the subcomplex $U \setminus \set{E,F}$ and add the cycle
$(p,E,q,F,p)$.
\end{enumerate}

The only unfolded complex that does not fall into one of these four
cases is a single simplex.
\end{proof}

Once we have an even scaffold, we can make it connected using local
flip operations as in the two-dimensional case; in fact the proof is
slightly simpler because every component of an even scaffold is
$2$-connected.  Let $A$ and $B$ be adjacent simplices that lie in
different components of the even scaffold, and suppose the scaffold
contains edges $(p,A)$, $(q,A)$, $(r,B)$, and $(s,B)$.  The ridge
$A\cap B$ contains all but one vertex of $A$ and all but one vertex of
$B$, so without loss of generality, $q$ and $r$ are both in $A\cap B$.
If we replace edges $(q,A)$ and $(r,B)$ with edges $(q,B)$ and
$(r,A)$, we obtain a new even scaffold.  Any node in the old component
of $A$ is still connected to $p$, then $A$, then $q$, and then to any
node in the old component of $B$.  Thus, the new even scaffold has one
fewer component.  Repeating this process for each adjacent pair of
components, we obtain a connected even scaffold.

Putting the pieces together yields the following:

\begin{theorem}
\theolab{d-manif.cycle} For any $d\ge 3$, every connected simplicial
$d$-manifold (possibly with boundary) has a facet cycle, except a
single $d$-simplex.
\end{theorem}

As we only need a path for the slab construction, we immediately
obtain:

\begin{cor}
Every connected simplicial manifold (possibly with boundary) has a
vertex-unfolding, which can be computed in linear time.
\end{cor}

%----------------------------------------------------------------------
\section{Discussion}
\seclab{Discussion}

The obvious question left open by our work is whether the restriction
to simplicial facets is necessary.  Does every three-dimensional
polyhedron with simply-connected facets have a non-overlapping
vertex-unfolding?  What if we require the facets to be convex?

Our strip construction fails for polyhedra with non-triangular convex
facets, because such polyhedra may not have facet paths.  For example,
the truncated cube has no facet path: no pair of its eight triangles
can be adjacent in a path, but its six octagons are not enough to
separate the triangles.

If facets are permitted to have holes, then there are polyhedra that
cannot be vertex-unfolded at all, for example, the box-on-top-of-a-box
construction of Biedl et al. (Figure~7 of~\cite{bddloorw-uscop-98}).

% cf. email from Jeff Erickson on June 22, 2001
A related reverse problem is, given a collection of polygons glued together at
vertices, can they be glued along their edges to form a polyhedron?
In other words, do they form a vertex-unfolding of some polyhedron?
What about convex polyhedra?  What is the complexity of these
decision problems?
%The latter problem is related to the still-open
%variant on \cite{Lubiw-O'Rourke-1996} in which the creases are prescribed.

\paragraph{Acknowledgments}
We thank Anna Lubiw for a clarifying discussion, and Allison Baird,
Dessislava Michaylova, and Amanda Toop for assisting with the
implementation.

\bibliographystyle{alpha}
%JOR's geom.bib file.  Use the .bbl file instead...
\bibliography{/home1/orourke/bib/geom/geom}

\end{document}